# ARTICLE

# Vibrational properties of graphdiynes as 2D carbon materials beyond graphene

P. Serafini,[a] A. Milani,[a] M. Tommasini,[b] C. Castiglioni,[b] D.M. Proserpio,[c] C.E. Bottani,[a] C.S. Casari*[a]

Two-dimensional (2D) hybrid sp-$sp^2$ carbon systems are an appealing subject for science and technology. For these materials, topology and structure significantly affect electronic and vibrational properties. We investigate here by periodic density-functional theory (DFT) calculations we here investigate the Raman and IR spectra of 2D carbon crystals belonging to the family of graphdiynes (GDYs) and having different structures and topologies. By joining DFT calculations with symmetry analysis, we assign the IR and Raman modes in the spectra of all the investigated systems. On this basis, we discuss how the modulation of the Raman and IR active bands depends on the different interactions between sp and $sp^2$ domains. The symmetry-based classification allows identifying the marker bands sensitive to the different peculiar topologies. These results show the effectiveness of vibrational spectroscopy for the characterization of new nanostructures, deepening the knowledge of the subtle interactions that take place in these 2D materials.

## 1 Introduction

In the last decades, carbon materials and their nanostructures played an outstanding role in science and technology. Fullerenes, graphene, carbon nanotubes, and other carbon-based polyconjugated polymers are fundamental milestones that paved the way to the so-called "era of carbon allotropes", enlightened by ground-breaking results and Nobel prizes [1].

In the last years, the interest has been directed toward novel forms of carbon, including 1D and 2D systems based on sp-hybridized carbon (variously referred to as carbyne, carbon atom wires, polyynes, cumulenes, ...) and on hybrid sp–$sp^2$ carbon systems. These investigations focused both on the fundamental properties and the potential applications in different fields, showing promising perspectives for the near future [2–6]. Graphyne (GY) and graphdiyne (GDY) represent 2D carbon crystals with sp–$sp^2$ carbon atoms [7–14]. They can be considered as a possible modification of graphene by interconnecting $sp^2$-carbon hexagons with linear sp-carbon chains of different lengths (a single or a double acetylenic bond for GY and GDY, respectively), generating new systems with peculiar and tunable electronic and optical properties. Starting from GY and GDY and playing with geometry and topology, many other 2D hybrid sp–$sp^2$ carbon 2D systems can be proposed, offering countless possibilities in the design and tailoring of carbon allotropes both theoretically and experimentally. Early theoretical studies on GY- and GDY-based systems were reported in 1987 [15] and more recently modern computational methods have been employed to shed light on their properties [16–20].

These structures can be classified as covalent organic frameworks (COFs), showing the occurrence of Dirac cones, flat bands, and tunable bandgap [21,22]. Such features of the electronic structure of COFs have been explained based on peculiar topological effects also in connection to their influence on the charge transport behaviour [23–26]. Several papers report on the prediction of properties of γ-GDY predicted mainly through density functional theory (DFT) calculations [5,6,27–32]. The thermal stability has been addressed by molecular dynamics simulations showing stability up to 1000 K for 6-R structure and up to 1500 K for α-, β- and γ-GDY when in the form of nanoscrolls (i.e. wrapped up sheets) [33,34].

Vibrational spectroscopy turned out to be extremely useful to characterize carbon-based materials. Raman spectroscopy has been employed in the first experimental works reporting the preparation of GDY-based materials [35–39]. Despite the effectiveness of these spectroscopic techniques up to now and to our knowledge, only a few investigations focused on the prediction of the vibrational properties of GDY [37–39]. Popov et al. computed the Raman spectrum of different polymorphs of GY [36], Zhang et al. gave an assignment of the Raman spectrum of GY and GDY and studied the effects of strain by Raman [37]. Wang et al. discussed the Raman spectra of different finite-size models of different sp-$sp^2$ carbon nanostructures [39] while IR active C≡C stretching band has been discussed in Ref. [38]. In a previous paper, some of us carried out a detailed vibrational analysis of γ-GDY crystal and its nanoribbons, including a discussion of the associated Raman and IR spectra [31]. Nevertheless, it is still lacking an explorative vibrational analysis spanning across different graphdiynes, other than γ-GDY.

From the experimental side, synthetic bottom-up approaches have been successfully employed by Haley and co-workers [40–47] to produce sub-fragments of GDY of different topologies and dimensions. Later, starting from the pioneering work done by Liliang Li and co-workers showing the route to the synthesis of graphdiyne materials [48], many papers reported the preparation and characterization of extended GDY sheets prepared through organometallic synthesis techniques [48–52]. The development of routes to the production of these systems allowed to employ GDY in different applications, even though significant efforts should still be put into further investigating and understanding their properties [52]. Recent advances in on-surface synthesis allow the preparation of hybrid sp–$sp^2$ carbon nanostructures and their atomic-scale investigation with surface science techniques [9,53–57]. Hence, the recent possibilities to realize new systems have opened new opportunities well beyond the sole investigation of their fundamental properties. In a previous paper, some of us developed an algorithm to systematically generate 2D carbon crystals belonging to the family of graphdiynes (GDYs) and having different structures and sp/$sp^2$ carbon ratios. We analyzed how structural and topological effects can tune the relative stability and the electronic behavior, to propose a rationale for the development of new systems with tailored properties [58].

Here, by using the set of 2D crystals generated in our previous work we investigate in detail the vibrational pattern of various graphdiyne systems. Two families of 2D materials have been considered. The first is composed of the simplest topologies of a single repeating unit, i.e. line (L), hexagon (H), and rhombus (R), and on widely studied polymorphs of GDY. The second family of 2D crystals (hereafter referred to as grazdiynes, GZY) contains linear diacetylene chains (L) and extended $sp^2$ domains forming zig-zag sequences of CC bonds like polyacetylene and/or ribbons made by condensed aromatic rings with L chains linked to the zig-zag edges (see [58] for details on the structural models and their construction). For both groups of materials, we report an investigation of their vibrational structure, unveiling how different physicochemical effects can influence the vibrational spectra of GDY and GZY systems. For the first family (GDYs), the effect of topology and stability on the main Raman active bands is correlated to different structural features of the sp and $sp^2$ domains. At the same time, the IR spectra have been investigated to identify marker bands able to uniquely characterize the different 2D crystals. The same analysis extended to the second family shows how a different relative number of aromatic and L units affects the vibrational response of these materials. In these cases (GZYs), the Raman and IR spectra were investigated to correlate the crystal structure with the vibrational structure and π-conjugation effects. Our findings are relevant for the spectroscopic characterization of these materials and may provide an insight that could lead to the design of new syntheses.

## 2 Computational details

To systematically identify all the sp–$sp^2$ carbon systems in the GDY family, we used ToposPro to generate subnets of graphene where bonds are deleted in all possible ways and substituted by linear diacetylenic units (see [58] for details). Based on the 2D crystal structures selected by ToposPro, periodic boundary condition (PBC) DFT simulations have been carried out by employing CRYSTAL17 to optimize the geometry (both the atomic position and cell parameters) and to compute the Raman and IR spectra. We adopted the hybrid exchange–correlation functional PBE0 together with the 6-31G(d) Gaussian basis set. This level of theory has been chosen according to our previous investigations of the structural and vibrational properties of γ-GDY and related nanoribbons, in which we compared the results obtained using different functionals and basis sets [31]. When using the 6-31G(d) basis set in PBC-DFT simulations with the CRYSTAL code, the exponent of the diffuse sp orbitals of carbon atoms has been increased from 0.1687144 to 0.187 $Bohr^{-2}$ to avoid convergence problems in the SCF [59], due to basis set linear dependencies. Considering the other simulation parameters, the tolerance on integral screening has been fixed to 9,9,9,9,80 (TOLINTEG parameters), whereas the shrink parameters defining Monkhorst–Pack and Gilat sampling points have been fixed to 50 for the prediction of the vibrational properties. We report unscaled wave-number values both in the discussion and in the figures (a suitable scaling factor should be defined whenever comparing these theoretical spectra with experimental ones, see [60]). To ease the comparison among the different crystals, we adopted the same Cartesian reference system with the z-axis orthogonal to the 2D crystal plane. When useful, a local reference system (x', y', z') is introduced for the description of physical quantities characteristic of the linear diacetylene branches: the x' axis is oriented along the sp chain and z' parallel to the z-axis of the crystal.

## 3 Results and Discussion

In our previous paper [58], we developed an approach to generate and classify all possible GDY/GZY 2D structures. We considered stable graphene derivatives by inserting linear diacetylene (C4) groups in the primitive cell. Our approach is based on removing edges from the graphene structure and substituting them with linear diacetylenic units. With the help of the topological classification tools in ToposPro, we extracted 332 topologically distinct patterns and then, by defining five building blocks, here named h (small hexagon), H (large hexagon), R (rhombus), T (triangle), L (line), we extracted only those containing possible building blocks that allow to tile the plane without a large distortion. By such procedure, we obtained 26 GDY-like structures, of which 17 are new. The structures are named as 6-$h^n L^m T^o R^p H^q$, where 6 represents the number of carbon atoms along the longest edge and the superscript on the building block symbols represents the number of each block appearing in the primitive unit cell. The theoretical analysis of novel carbonaceous materials has been already performed by substituting atoms in graphene layers or fragments (see e.g. [61,62]). Such approaches however modeled doping of carbon by heteroatoms and focused on the formation of local defects and their effects on the electronic properties. Our approach was focused on the identification of novel extended 2D crystals based on a systematic generation and selection of possible GDY structures generated from graphene by the introduction of diacetylene units [58].

### 3.1 Electronic features and molecular structure

We focus first on 6-L, 6-H (also known as α-GDY [15, 16]), and 6-R, which are characterized by the simplest topologies of a single repeating unit (line, large hexagon (H), and rhombus (R), and on 6-hT$^2$ and 6-HT$^2$ which are the two widely studied polymorphs of graphdiyne also called γ-GDY and β-GDY (respectively) [15, 16]. A second family of 2D crystals (hereafter referred to as grazdiynes, GZY) contains linear diacetylene chains (L) and extended sp$^2$ domains forming zig-zag sequences of CC bonds like polyacetylene and/or ribbons made by condensed aromatic rings with L chains linked to the zig-zag edges. All these systems are sketched in Figure 1.

The electronic structures of the 2D crystals under investigation display some common characteristics. A simplified description of the crystal orbitals, following the theory of molecular orbitals as a linear combination of atomic orbitals (MO=LCAO) and applied to the minimal set consisting of the hybrid atomic orbitals (AO) belonging to the valence shell of the carbon atom, gives us some useful insights. In particular, it highlights the possible role and interactions of sp$^2$-C vs sp-C in determining the molecular architecture (e.g., bond lengths) as well as physical properties (e.g., the spectroscopic response). The orbitals belonging to the highest energy occupied bands and to the lowest energy unoccupied bands should mainly arise by the combination of p$_z$ orbitals of sp and sp$^2$ hybridized carbon atoms and by the combination of the p$_{y'}$ orbitals of the sp-carbon chains. Moreover, because of the mirror symmetry to the crystal plane, π$_z$ orbitals coming by the mixing of p$_z$ cannot mix with σ orbitals from any combinations of p$_{y'}$ orbitals.

We underline that: (i) The different crystal structures and the topology of the bond network rule the features of the π$_z$ orbital, which could involve the whole set of p$_z$ AOs (with similar weights of the p$_z$ AO from sp$^2$- and sp-carbon atoms) or could be preferentially localized either on sp$^2$- or on sp-carbon atoms. According to the above property of π$_z$ orbitals, the resulting electron charge density is fully delocalized over the whole crystal network or is mainly localized either on the individual diacetylene segments or sp$^2$ domains. (ii) Since the p$_{y'}$ AOs of sp carbons possess even mirror symmetry to the xy plane, they can mix with the sp$^2$-hybridized AOs of the sp$^2$-carbon atoms (same even symmetry), and therefore may participate to the σ orbitals system. However, based on energy considerations and local symmetry (they are odd to the vertical mirror plane x'z' whereas the sp$^2$ hybrids are even), this mix is negligible and p$_{y'}$ AO should concur to the formation of well-localized MO on the individual diacetylene segments. Electrons sitting in these orbitals impart the quasi-triple bond character exhibited by the two bonds adjacent to the central diacetylene quasi single CC bond (P$_i$ label in Fig. 2). (iii) Considering point i) each of the crystals presented in Fig. 1 possesses an extended network of p$_z$ electrons (one p$_z$ AO at every C site), which possibly give rise to highly delocalized electron density. On the other hand, the delocalization properties are modulated by the interplay between the sp and the sp$^2$ phase, which depends on the selected crystal structure. This feature strongly affects t.he degree of π- conjugation among different diacetylene chains.

Since the vibrational spectrum (notably Raman) of sp carbon chains is very sensitive to the delocalization properties of the π electrons, as well as to end group effects [35], we analyse the computed spectra of GDYs and GZYs, focusing mainly on the vibrational transitions involving the diacetylene branches.

Table 1. For each structure, the energy contribution per carbon atom referred to graphene (ΔE) and bond alternation parameter (BLA) of the polyyne units are reported in the table, together with the number and symmetry species of polyyne ECC phonons.

| structure | ΔE (kcal mol$^{-1}$) | BLA (Å) | ECC modes |
|---|---|---|---|
| 6-H (α -GDY) | 27.39 | 0.142 | 1A$_g$+1E$_{2g}$ |
| 6-L | 26.18 | 0.136 | 1A$_g$ |
| 6-R | 25.48 | 0.154 | 1A$_g$+1B$_{1g}$ |
| 6-HT$^2$ (β -GDY) | 25.40 | 0.156 | 1A$_g$+2E$_{2g}$ |
| 6-hT$^2$ (γ -GDY) | 21.13 | 0.167 | 1A$_g$+1E$_{2g}$ |
| 6-hL | 20.59 | 0.138 | 1A$_g$ |
| 6-hL$^2$ | 23.11 | 0.136 | 2A$_g$ |
| 6-hL$^3$ | 24.06 | 0.135 | 3A$_g$ |
| 6-h$^2$L | 16.81 | 0.139 | 1A$_g$ |
| 6-h$^3$L | 14.15 | 0.138 | 1A$_g$ |

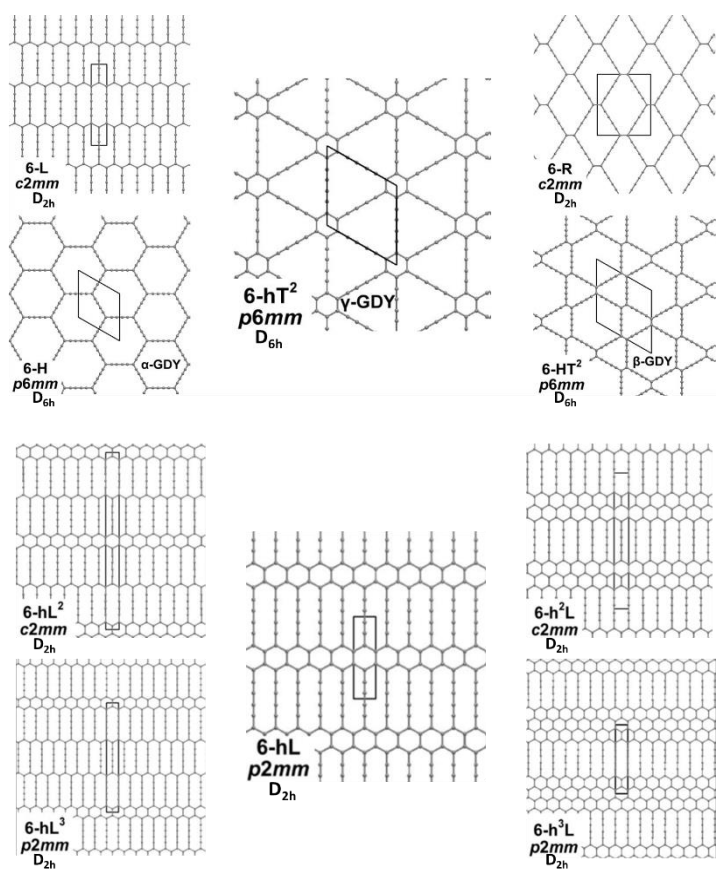

Figure 1: Sketch of the 2D-GDY (top) and 2D-GZY (bottom) structures analysed. For each structure, the unit cell, plane group and point group symmetry (factor group at Γ point of BZ) are illustrated.

### 3.2 Vibrational spectra of GDY

We start by defining the internal and symmetry CC stretching coordinates of the diacetylene branches. This allows establishing the peculiar group vibrations of the chains that are involved in the Raman and IR active q=0 phonons, with no need for calculation. For each diacetylene branch (labeled by the i index) we consider the sequence of the five CC bonds forming the 6 carbon atoms chain, and we define the relative stretching coordinates ($r_i$, $r'_i$, $T_i$, $T'_i$, $P_i$) as indicated in Figure 2. We also define the group coordinates {$S_i$} sketched in Figure 2: the symmetric ($T_i^+$) and the antisymmetric ($T_i^-$) combination of quasi-triple CC bonds stretching; the symmetric ($r_i^+$) and the antisymmetric ($r_i^-$) combination of the terminal quasi-single CC bonds stretching, and the stretching of the central quasi-single CC bond ($P_i$). The group coordinates defined above are symmetry coordinates for the diacetylene fragment, showing an inversion centre in the middle of the central CC bond.

In most of the crystal structures illustrated in Fig. 1 and Table S1 in SI, the inversion centres carried by each diacetylene unit are preserved, hence the group coordinates {$S_i$} define a symmetry-adapted basis set of internal coordinates that is useful for the interpretation of the spectral features of the crystals. This happens for the crystals 6-H, 6-L, 6-R and 6-hT$^2$, whereas in 6-HT$^2$ the inversion centres located on the chains are lost. Remarkably, for all the crystals but 6-HT$^2$, we can state that within the {$S_i$} set only the subset of gerade coordinates {$S_k$}$^g$ = {$r_i^+$, $T_i^+$, $P_i$} is involved in the Raman active phonons, while the subset of ungerade coordinates {$S_k$}$^u$ ={$r_i^-$,$T_i^-$} is involved in the IR active phonons. This feature is independent on the number of diacetylene chains in the unit cell and on the kind of chemical group linking the chains. On other hand, the symmetry of the crystal and the number of chains will determine the suitable symmetry combinations of the {$S_k$}$^g$ of the different chains, giving Raman active phonons. The $A_{1g}$ and $E_{2g}$ linear combinations of the {$S_k$}$^g$ coordinates of different chains describe the Raman active phonons of the crystals possessing $D_{6h}$ symmetry (6-H, and 6-hT$^2$). In the 6-L and 6-R structures ($D_{2h}$ symmetry), such linear combinations give rise to $B_{1g}$ and $A_g$ Raman active phonons. Similarly, linear combinations of {$S_k$}$^u$ coordinates give $E_{1u}$ IR-active phonons in $D_{6h}$ structures and $B_{3u}$/$B_{2u}$ IR-active phonons in $D_{2h}$ structures.

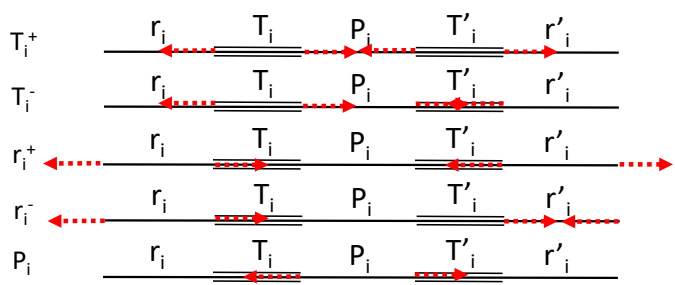

Figure 2. Definition of the CC stretching internal coordinates and sketches of their symmetry adapted combinations {Sk}.

Vibrational displacements located in the sp$^2$ domains will also concur with the definition of the vibrational trajectories associated with the phonons, giving rise to rather complex patterns of collective atomic displacements. Luckily, the definition of the sets {$S_k$}$^g$ and {$S_k$}$^u$ greatly eases the analysis of the features of the informative spectral region above 2000 cm$^{-1}$ that mainly involves vibrations of the diacetylene branches.

In Figure 3 we report the computed Raman and IR spectra of the investigated series of GDY. The frequencies of the optical phonons and the predicted Raman and IR band intensities are shown in Table S2, together with the classification of the phonons according to the suitable point group symmetry (namely the factor point symmetry group at the Γ point of the first Brillouin Zone BZ). In Table 2 we decided to report only those phonons with high vibrational frequency (> 2000 cm$^{-1}$) that are related to stretching modes of the diacetylene chains.

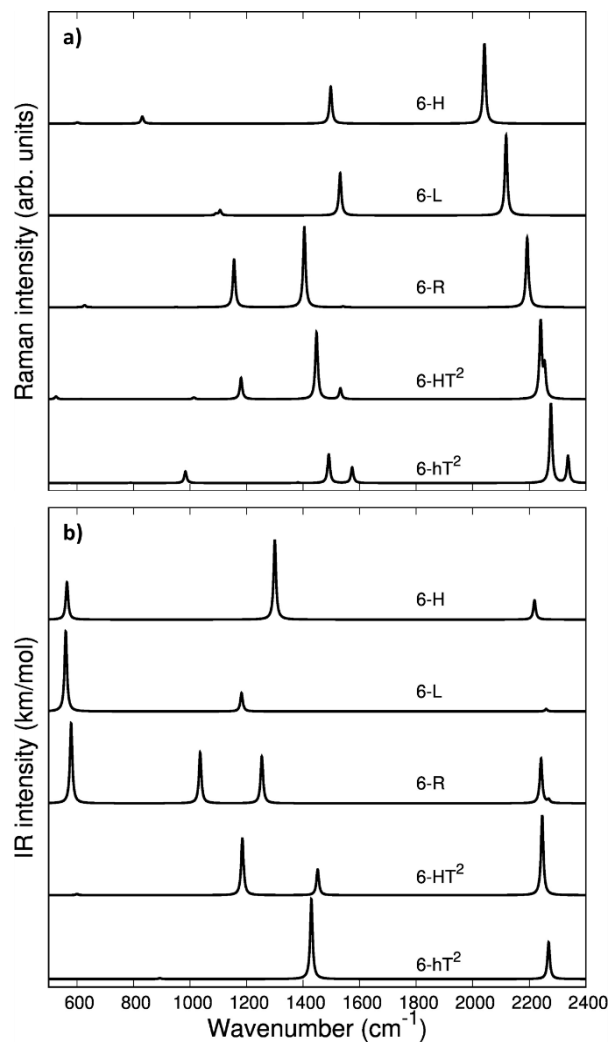

Figure 3. Computed Raman (a) and IR (b) spectra of GDY systems: 6-hT$^2$, 6-HT$^2$, 6-R, 6-L and 6-H.

The vibrational spectra of GDYs, show two characteristic regions. The 2000-2400 cm$^{-1}$ interval is typical of the stretching modes of the CC triple bonds. The vibrations of the sp-carbon chains give rise to strong Raman active bands above 2000 cm$^{-1}$, which can be considered the marker bands of the diacetylenic units. In the lower frequency region (800-1600 cm$^{-1}$), we find Raman active bands assigned to the stretching vibrations of the CC bonds linking sp$^2$ hybridized carbon atoms, *i.e.*, bonds belonging to aromatic rings (case 6-hT$^2$) or the CC bonds linking the diacetylene branches (cases 6-L, 6-R, 6-HT$^2$). Such phonons always involve additional contributions from the stretching of the quasi-single bonds belonging to the sp-carbon chains.

Table 2. Computed wavenumbers (spectral region above 1800 cm$^{-1}$), symmetry species, relative Raman activities and IR band intensities for the **q=0** Raman or IR active phonons of the GDY and GZY crystals. Relative Raman activities are calculated by assigning the value of 1000 to the strongest Raman transition of the whole spectrum.

| Crystal | Raman Spectrum | | | IR spectrum | |
|---|---|---|---|---|---|
| | wavenumber (cm$^{-1}$) (Irrep) | Raman activity | Vibrational assignment | wavenumber (cm$^{-1}$) (Irrep) | IR intensity (km mol$^{-1}$) |
| **2D- GDY structures** | | | | | |
| **6-H** | 2041 ($E_{2g}$) | 1000 | ECC | 2219 ($E_{1u}$) | 365 |
| | 2361 ($A_{1g}$) | 0.02 | ECC | | |
| **6-L** | 2118 ($A_g$) | 1000 | ECC | 2260 ($B_{3u}$) | 2 |
| **6-R** | 2193 ($A_g$) | 1000 | ECC | 2241 ($B_{3u}$) | 267 |
| | 2202 ($B_{1g}$) | 57 | ECC | 2269 ($B_{2u}$) | 20 |
| **6-HT²** | 2240 ($E_{2g}$) | 142 | partial ECC character | 2246 ($E_{1u}$) | 1501 |
| | 2240 ($A_{1g}$) | 1000 | ECC | 2356 ($E_{1u}$) | 2 |
| | 2255 ($E_{2g}$) | 453 | partial ECC character | | |
| **6-hT²** | 2277 ($A_{1g}$) | 1000 | ECC | 2269 ($E_{1u}$) | 480 |
| | 2337 ($E_{2g}$) | 344 | ECC | | |
| **2D-GZY structures** | | | | | |
| Crystal | Wavenumber (cm$^{-1}$) (Irrep) | Raman activity | Vibrational assignment | Wavenumber (cm$^{-1}$) (Irrep) | IR intensity (km mol$^{-1}$) |
| **6-hL** | 2248 ($A_g$) | 328 | ECC | 2255 ($B_{3u}$) | 8719 |
| **6-hL²** | 2205 ($A_g$) | 1000 | partial ECC character | 2255 ($B_{3u}$) | 584 |
| | 2244 ($A_g$) | 358 | partial ECC character | 2310 ($B_{3u}$) | 58 |
| **6-hL³** | 2173 ($A_g$) | 1000 | ECC | 2219 ($B_{3u}$) | 444 |
| | 2245 ($A_g$) | 107 | C≡C str | 2256 ($B_{3u}$) | 248804 |
| | 2313 ($A_g$) | 15 | ECC | 2292 ($B_{3u}$) | 88 |
| **6-h²L** | 2257 ($A_g$) | 1000 | ECC | 2250 ($B_{3u}$) | 640 |
| **6-h³L** | 2224 ($A_g$) | 387 | ECC | 2247 ($B_{3u}$) | 512 |

**Vibrational transitions in the 2000 -2400 cm$^{-1}$ range.** The higher frequency region of the Raman spectrum of GDYs shows the strongest transitions in the whole Raman spectrum. According to our symmetry analysis, for $D_{6h}$ structures (6-H and 6-hT²) we can predict two Raman active phonons resulting from the $A_{1g}$ and $E_{2g}$ (doubly degenerate) combination of the symmetric triple bond stretching ($T_i^+$) of the three diacetylene chains belonging to the hexagonal cell. . For the 6-R and 6-L crystals we have find two $A_g$ and $B_{1g}$ transitions and just one $A_g$ transition, respectively. This follows the number of chains in the primitive cell of 6-R and 6-L (two and one chain, respectively).

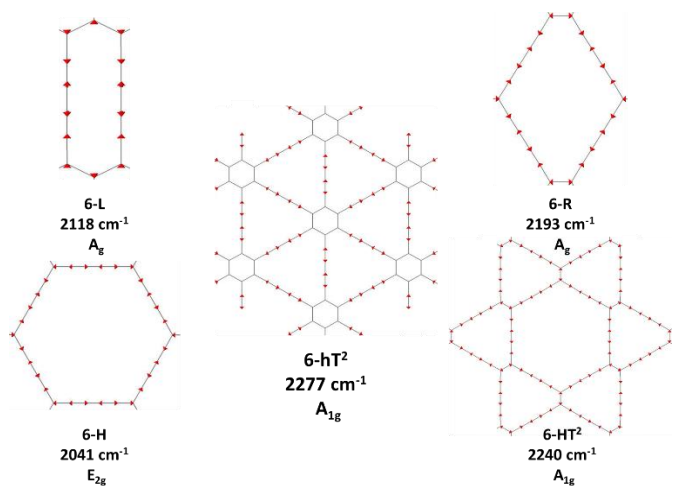

Figure 4. Sketches of the vibrational eigenvectors associated to the strongest Raman active ECC phonons of GDYs. The corresponding vibrational wavenumbers and symmetry species are reported.

The 6-HT$^2$ crystal is peculiar, with a cell containing six diacetylene branches, which lost their local inversion centre in the crystal. In this case, we obtain one $A_{1g}$ and two $E_{2g}$ bands, with the $E_{2g}$ transition at 2240 cm$^{-1}$ and the $A_{1g}$ transitions showing accidental degeneracy. In 6-HT$^2$ the two $E_{2g}$ phonons mix $T_i^+$ and $T_j^-$ group coordinates of different chains, while the $A_{1g}$ phonon corresponds to the in-phase combination of the six $T_i^+$, similar to the other GDYs (see Figure 4 and SI). With the only exception of the $E_{2g}$ vibrations of 6-HT$^2$, the Raman active phonons in the 2000-2400 cm$^{-1}$ region show, for all the diacetylene branches, the characteristic pattern of the Effective Conjugation Coordinate (ECC) mode. In our case, the ECC is a collective simultaneous C≡C stretching/C–C shrinking of the bonds forming the diyne segments (see sketches of the vibrational eigenvectors in Figure 4 and SI). Indeed, the group coordinates $T_i^+$ are dynamically coupled with the other stretching degrees of freedom of the sp chain, thus giving rise to collective vibrations, involving also stretching vibrations of the quasi single CC bonds, namely of the sub-sets of coordinates {$r_i^+$ , $P_i$}, which oscillate out-of-phase with $T_i^+$, during the ECC vibration. The remarkable dynamic coupling between the triple and single bond stretching in the sp-carbon chain, resulting in ECC-modes, is a well-recognized phenomenon, encountered for a wide variety of oligoynes, irrespective of their chain length and end groups [35, 63-66]. The high Raman activity of the ECC modes and the sensitivity of their frequency to π-electron conjugation have been widely investigated in the past and are reported as the signature of the strong electron-phonon interaction that takes place along the ECC vibrational trajectory [63,65,66]. The ECC vibration can be described as the oscillation of the bond length alternation (BLA) parameter, which is the key vibrational parameter affecting the electronic structure of carbon chains with polyconjugated π electrons [63,65,66].

Interestingly, in GDY crystals, the occurrence of a shrinking of the C-C bonds of the sp chain that is simultaneous with the C≡C stretching not only depends on the dynamical coupling but is a direct consequence of the translational symmetry of the crystal. It requires that q=0 phonons are such that the atomic displacements do not modify the cell volume, at any time. Indeed, during ECC vibrations, the shrinking of the C-C bonds (implied by contributions from $P_i$ and $r_i^+$) does guarantee the fulfillment of such translational symmetry constraint when the two C≡C bonds of diacetylene segment stretch in phase ($T_i^+$).

Looking at Figure 3, we observe for the 6-hT$^2$ system two distinct bands at rather different frequencies associated respectively to the $A_{1g}$ symmetry ECC phonon (2277 cm$^{-1}$) and to the $E_{2g}$ ECC phonon (weaker band at 2337 cm$^{-1}$). In a previous paper [31] we showed that the sizeable difference in wavenumber of these two phonons (60 cm$^{-1}$) is due to the interaction among adjacent diacetylene chains, linked through the CC bond belonging to the phenyl unit. The separation in frequency between such ECC modes is smaller in other GDYs, whereas in the most cases the intensity of the totally symmetric $A_{1g}$ (or $A_g$) line is the largest one. However, 6-H represents a peculiar exception in the GDY family: in this case, the ECC $E_{2g}$ phonon has the largest Raman activity, while the $A_{1g}$ phonon is practically silent (it is 10$^{-5}$ times weaker than the $E_{2g}$ ECC phonon -see Table 2 and S2). Interestingly, for 6-H the frequency of the $E_{2g}$ phonon is very low (2041 cm$^{-1}$), and remarkably lower than that of the $A_{1g}$ mode, (2361 cm$^{-1}$) which is up-shifted by 320 cm$^{-1}$. Such features indicate that, differently from the other GDYs, the effect of the interactions between diacetylene chains coupled through just one sp$^2$ carbon atom cannot be described anymore as a weak perturbation of the physics of the individual sp chains. In other words, the vibrational dynamics and the Raman response of 6-H is heavily affected by the extended network of conjugated π$^z$-electrons involving all the diacetylene branches, linked by the single sp$^2$ carbon atoms that support the threefold branching.

As a result of the different behaviours commented above, the higher frequency region of the GDY series shows a very simple Raman pattern, with a strong ECC line, which is accompanied in 6-hT$^2$ and 6-HT$^2$ by only one weaker satellite band, located at a higher frequency (see Figure 3).

Figure 3 shows that the main ECC line follows a decreasing frequency trend as a function of the crystal structure, spanning a wide interval of about 300 cm$^{-1}$, along the sequence: 6-hT$^2$, 6-HT$^2$, 6-R, 6-L and 6-H (from the highest to the lowest $\nu_{ECC}$). This decreasing trend of $\nu_{ECC}$ parallels the increasing π-conjugation of the diacetylene chains, as observed elsewhere [31,32,35,63, 65,66]. This frequency behaviour is commonly reported in polyconjugated sp chain molecules [35, 63,65,66] in conjunction with the equalization of the equilibrium bond lengths (i.e., small BLA values). A similar behavior of decreasing BLA values with increasing π-conjugation is observed also in the present case (see Table 1 and Table S1).

In [58], a comparative study of the energy of several GDY crystal is reported. The crystal energy, normalized to the number of C atoms, is referred to the case of graphene, showing that the introduction of diacetylene units has an energy cost (positive ΔE, see values in Table 1 and S1). Moreover, ΔE values for the GDY series show a marked dependence on the crystal structure, e. g. on the relative abundance of sp$^2$ and sp phase and bonds topology: the most stable system is 6-hT$^2$ (γ-GDY) whereas 6-H (α-GDY) is the system with highest energy.

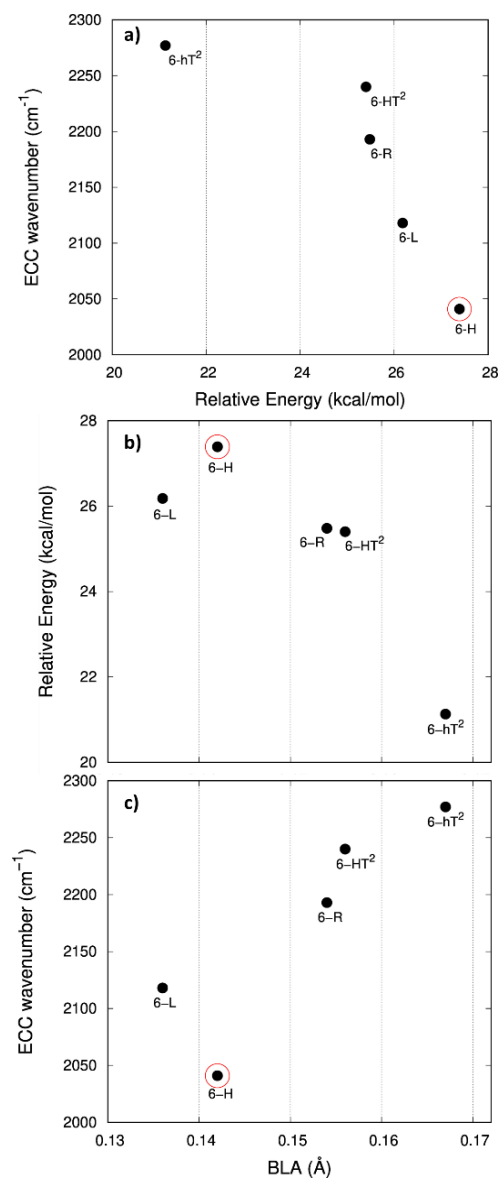

Figure 5. Correlation between physical quantities sensitive to π-electrons conjugation. Panel a: ECC phonon wavenumber *vs.* ΔE (ΔE is the relative crystal energy - per atom -referred to graphene, see text). Panel b: ΔE vs. BLA parameter. Panel c: ECC phonon wavenumber *vs* BLA. The data corresponding to 6-H crystal is highlighted by a red circle.

In Figure 5.a we report the plot of $\nu_{ECC}$ vs ΔE, which shows a nice correlation: $\nu_{ECC}$ softens as ΔE increases. By considering the high reactivity of the sp carbon chains characterized by a cumulene-like structure (*i.e.*, an sp chain ideally approaching a linear sequence of double bonds), we expect that the systems showing higher ΔE values also display a chain structure with more equalized CC bonds. The BLA parameter, which measures the average degree of bond alternation in the sp chains, provides an estimate of this property. The BLA parameter has been obtained as the difference between the average bond length of the quasi-single bonds and the average bond length of the quasi-triple bonds:

BLA = 1/3 ($r_0$+$r_0$'+$P_0$) - ½ ($T_0$+$T_0$')

where $r_0$, $P_0$, and $T_0$ are the equilibrium bond length values of the CC bonds belonging to the diacetylene chains (see Figure 2 for the meaning of the symbols).

The BLA values of GDYs, reported in Table 1, are compared to ΔE in the plot in Figure 5.b, showing the expected correlation. This plot nicely explains the observed correlation between energy and ECC frequency: as a matter of fact, the usual trend among $\nu_{ECC}$ and BLA values is indeed verified (Figure 5.c).

A closer look at Figure 5, shows that there is a frequency value, namely that of the 6-H crystal, which does not follow the approximately linear trend of $\nu_{ECC}$ with BLA. Indeed, 6-H shows an ECC frequency remarkably low in comparison with its BLA. Moreover, also in the plot of ΔE vs. BLA (Figure 5.b), the 6-H crystal is not within the trend. It seems that the predicted chain geometry does not fully reflect the very high electron delocalization described by $\nu_{ECC}$ and ΔE. Notice moreover that the 6-H structure also shows a peculiar vibrational feature, namely a dominant ECC band of $E_{2g}$ symmetry, which would deserve further investigations.

We tried to rationalize the origin of the different degrees of electrons conjugation in GDYs, by considering the different characteristics of the $sp^2$ linkers. As reported in [31], the high aromatic character of the phenyl units prevents an effective conjugation of the $\pi_z$-system of the ring with the $\pi_z$-electrons of the grafted diacetylene segments, which results in the large BLA and rather high frequency of the ECC Raman band of 6-$hT^2$. Instead, the sp chains belonging to 6-$HT^2$ and 6-R, are linked to each other through just one CC bond formed between two $sp^2$ C atoms (i.e., through a formal double bond). This kind of linker is more prone to share its $\pi_z$ electrons with the $\pi_z$-electron system of the diacetylene units. Thus, 6-$HT^2$ and 6-R show intermediate ECC frequency and BLA values. Also in 6-L, which features arrays of $sp^2$ C atoms forming zig-zag chains with equalized CC bond length (a polyacetylene-like chain) one can identify pathways consisting of annfinite sequence of sp chains bridged by one CC bond formed by $sp^2$ C atoms. In this case, the $p_z$ electrons of the $sp^2$ C atoms belong to a fully delocalized π system along the zig-zag chain which is more effective in promoting $\pi^z$-electron conjugation with the sp branches. The 6-L structure is in the second place if one considers the trend of ΔE and ECC frequency (from lower to higher frequency values). A further enhanced tendency toward π-conjugation is observed in 6-H, with just one $sp^2$ C atom acting as a node of the network of diacetylene chain, showing the lower ECC frequency, in correspondence with the higher instability of the crystal (higher ΔE).

For most GDYs $\nu_{ECC}$ shifts toward lower values as the ratio between the number of sp and $sp^2$ C atoms (sp/$sp^2$) decreases. As observed in [58], the sp/$sp^2$ ratio also correlates with the relative energy, showing a general increase with the increase of the sp phase content. 6-L is the only exception to this trend and it possesses a very low $\nu_{ECC}$ and a high ΔE value but, in contrast, it features a low sp/$sp^2$ carbon ratio: this observation tell us that also the topology of the $sp^2$ phase can play a very important role.

The infrared spectra of the higher frequency region is characterized by phonons involving the antisymmetric stretchings of the triple bonds ($T_i^-$) accompanied by out of phase contributions of ($r_i^-$). Because of the inversion symmetry, IR active phonons cannot involve the stretching of the central CC bonds.

In the $D_{6h}$ systems featuring three sp branches in the cell (6-H, 6-hT$^2$) we find only one IR active (doubly degenerate) $E_{1u}$ phonon and we find one more $E_{1u}$ phonon in the case of 6-HT$^2$ (with six sp branches). 6-L shows just one $B_{3u}$ transition and 6-R two ($B_{3u}$ and $B_{2u}$) transitions, according to the number of branches belonging to the primitive cell. Looking at the IR intensity values (Table 2), it is immediate to realize that, also in presence of two IR active modes, only one band is strong enough to be identified in the plot of the simulated IR spectra.

In the framework of ECC theory [35, 58], such IR active modes have been described as the BLA oscillation with a node in the middle of the chain. Differently from the collective ECC modes, the vibrational frequency of these modes is not much sensitive to the degree of conjugation of the chain, especially in the case of short sp chains, moreover their frequency is always higher than $\nu_{ECC}$. The above characteristics reflect also in the spectral trend, showing a modest dependence of the position of the high-frequency IR bands on the different crystal structures of the GDY series.

**Vibrational transitions below 1600 cm$^{-1}$.** The Raman transitions in this region correspond to phonons involving the stretching of the "quasi" single bonds of the sp chains. For a given i-th sp chain, the $r_i^+$ and $P_i$ coordinates are involved in out-of-phase combination ($r_i^+$-$P_i$), which fulfills the requirement that the cell volume is preserved during the vibrational displacements. The stronger Raman line in this region has a frequency value in the 1550 – 1480 cm$^{-1}$ range (see Tables S2 and S3) and belongs to the totally symmetric species for all the crystals. However, for 6-H, which shows a strong $E_{2g}$ ECC transition, also the $r_i^+$- $P_i$ phonon belongs to $E_{2g}$.

Table 3. The bond length of the quasi single CC bonds belonging to the diacetylene branches, for GDY structures. For the meaning of bond labels, see Figure 2.

| GDY | 6-R | 6-HT$^2$ | 6-hT$^2$ | 6-H | 6-L |
|---|---|---|---|---|---|
| P (Å) | 1.343 | 1.345 | 1.351 | 1.336 | 1.329 |
| r (Å) | 1.398 | 1.399 | 1.406 | 1.391 | 1.381 |
| $\nu$ (C-C) (cm$^{-1}$) | 1404 | 1448 | 1491 | 1498 | 1531 |

Table 3 shows the nice correlation between the strength of the quasi-single CC bonds belonging to the sp chain and the frequencies of the associated CC stretching phonons for the series 6-R, 6-H, 6-R: the shorter the bonds, the higher is the frequency. 6-HT$^2$ and 6-hT$^2$ do not follow this trend, likely because of the coupling with ring breathing vibration in 6-hT$^2$ and because of the more complex vibrational structure of 6-HT$^2$ (the crystal with the largest unit cell). However, the vibrational eigenvectors become much more complex in presence of bonds linking sp$^2$ carbon atoms, which are possibly subjected to strong dynamical coupling with $r_i$ stretchings. Also, vibrational modes rather localized in the sp$^2$ domain can occur, as in the case of 6-hT$^2$, for which the nature of these vibrations has been discussed in our previous paper [31]. In particular, the three bands located at 1382, 1491, and 1574 cm$^{-1}$ show large contributions by stretching vibrations of the phenyl units, as reported in Table S3 of the SI.

The collective character of the phonons of GDYs below 1400 cm$^{-1}$ makes it difficult to propose a comparative discussion. We limit ourselves to the observation that 6-R and 6-HT$^2$, both featuring isolated sp$^2$ CC bonds between diacetylene chains, present a strong band close to 1100 cm$^{-1}$, whose associated nuclear displacements show a remarkable contribution from the linker sp$^2$ CC bond (see Table S3).

Turning to the analysis of the IR spectra in the region below 1600 cm$^{-1}$, we focus our attention on the possibility to identify different marker bands for the different 2-D structures. For 6-H one can find a very intense IR band at 1300 cm$^{-1}$. Also for 6-hT$^2$, the peak with the highest IR intensity is located below 1500 cm$^{-1}$: we compute a strong band at 1429 cm$^{-1}$ with contributions from stretching vibrations of the bonds of the phenyl units (see Table S3). The marked frequency difference between such IR bands of 6-H and 6-hT$^2$ ($\Delta\nu$ = 129 cm$^{-1}$) makes it possible to spectroscopically distinguish the two structures.

The IR spectra of both 6-HT$^2$ and 6-R show two strong IR-active bands, well separated in frequency and located at different positions for the two systems. Looking at the vibrational eigenvectors, we can describe the associated vibrations as mainly characterized by in-phase $T_i^-$ and $r_i^-$. Also for these two GDYs, IR spectroscopy shows strong distinctive features that are useful for their identification.

6-L shows an isolated band located at 1182 cm$^{-1}$, very close to that of 6-HT$^2$ (1185 cm$^{-1}$). However, for 6-HT$^2$, the presence of the additional strong band of 6-HT$^2$ at 1452 cm$^{-1}$ allows distinguishing among the two crystals. Moreover, the whole intensity pattern is different for the two structures.

In the lower frequency region (< 600 cm$^{-1}$, see Figure 3) it is possible to identify different marker bands. Indeed, a very intense IR active peak appears for 6-R, 6-L and 6-H, located respectively at 580 cm$^{-1}$, 560 cm$^{-1}$ and 565 cm$^{-1}$. Through the analysis of the vibrational eigenvectors associated with these bands (SI) these phonons are characterized by collective in-plane bending of the CC bonds of the sp chains.

### 3.3 Vibrational spectra of GZY

The second class of crystals studied in this work consists in the series also known in the literature as grazynes [67]. They are 2D materials made by condensed aromatic rings forming ribbons with different widths connected by L units (diacetylene chains) (Figure 1) [67]. For the analysis of the Raman and IR spectra of GZY we will focus our attention on the evolution of the spectra inside two different families, whose structures are depicted in Figure 1.

The first family (GZY-I: 6-hL, 6-hL$^2$, 6-hL$^3$) shows a topology made by ribbons consisting of an infinite sequence of condensed h units (only one h unit in the height, forming a polyacene chain), intercalated by layers formed by L units. These layers have different heights, depending on the number n of L units (n =1,2,3), running parallel to the layer width (x-axis direction) and linked to each other through zig-zag polyacetylene-like sp$^2$ chains. The 6-L crystal, already discussed as a member of the GDY group, is a limiting case of the GZY-I family and will be considered in our comparative analysis.

The second family (GYZ-II: 6-hL, 6-h$^2$L, 6-h$^3$L) is characterized by layers of parallel L segments (the thickness of each layer corresponds to just one L unit) intercalated by nanoribbons formed by condensed h units with different widths. Graphene (6-h$^\infty$L) can be considered a limiting case belonging to this family. In Figure 6 we report the Raman spectra and the infrared spectra calculated for the GZY-I and GZY-II groups.

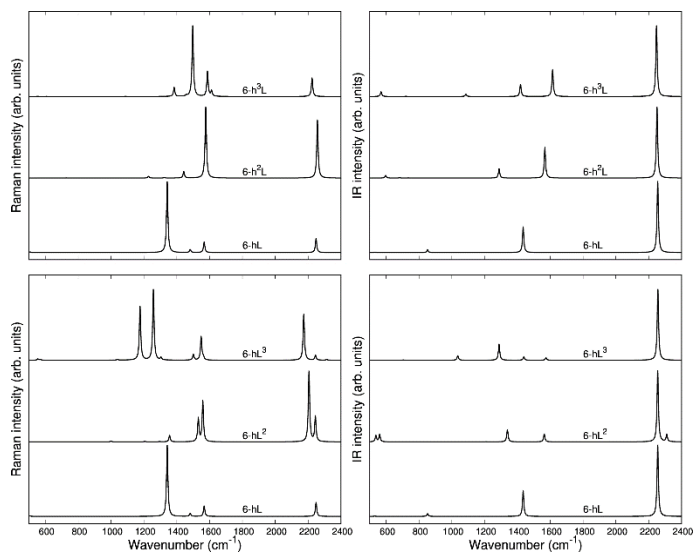

Figure 6. Calculated Raman spectra (left) and IR (right) spectra of GZY-I (top) and GZY-II (bottom).

The Raman spectra of GZYs show features typical of the sp phase, which is the main responsible for the characteristic vibrational features in the high frequency (above 2000 cm$^{-1}$) and of the graphene-like ribbons, which give rise to features below 1600 cm$^{-1}$. The interplay between the sp and sp$^2$ phases influences the degree of $\pi$ electrons delocalization in the whole crystal and is expected to have a signature, especially in the Raman spectra, which cannot be regarded by a mere superposition of the spectroscopic response of the L layers and the h ribbons. Because of the complex trajectories of the phonons below 1600 cm$^{-1}$, involving atomic displacements in both sp and sp$^2$ phase, the following discussion will be focused on the vibrational features coming from vibrations mainly localized on the diacetylene branches, which are found in the 2000 – 2400 cm$^{-1}$ region.

Raman spectra. The q = 0 phonons of GZY-I are classified according to the D$_{2h}$ symmetry: Raman transitions in the high-frequency range are associated with A$_g$ phonons. Table 2 lists the frequencies and intensities of these lines, which are as many as the number of sp chains belonging to the primitive cell. As in the case of GDY, one strong Raman line, possibly accompanied by satellite lines located on the higher frequency side, dominates the sp spectral region (Figure 6).

The eigenvectors analysis (Table 4) shows that only chains which carry a symmetry inversion centre develop a collective ECC vibration, namely the symmetric stretching of quasi triple bonds (T$_i^+$) accompanied by out-of-phase vibration of the quasi single bonds (r$_i^+$, R$_i$). This happens for 6-hL, and the central chains of the L layer of 6-hL$^3$. Chains that have lost the inversion symmetry give rise to phonons characterized by displacements localized either on the triple bond close to the zig-zag chain or close to the polyacene ribbon. This is the case of 6-hL$^2$: its strongest Raman active transition corresponds to a phonon describing a BLA oscillation along a path crossing a CC bond belonging to the zig-zag chain and does not involve the peripheral C≡C bond of the L$^2$ layer. Its low frequency and high intensity should be ascribed to delocalization of the $\pi_z$ electrons phenomena, which are coupled to the CC bonds of the polyacetylene-like chain. Similarly, we can justify the large Raman intensity (and low frequency) of the ECC-like mode of 6-hL$^3$. It can be described as due to the ECC vibration of the central chain (showing inversion centre) accompanied by the in-phase BLA oscillation of the two adjacent chains, localized on the C-C≡C sequence close to the zig-zig chain.

In the GZY-II family, the situation is simpler because all these structures contain only one L segment in the primitive unit cell, which is characterized by inversion centre. Thus, this family shows only one A$_g$ transition above 2000 cm$^{-1}$, corresponding to the ECC mode of the diacetylene chain. This implies only one Raman line in the higher frequency region of the Raman spectrum.

Interestingly, by looking at the calculated Raman spectra we can see that the GZY-II series shows a little dispersion of the ECC line, whereas for the GZY-I series the ECC – like phonon shows a dispersion of 130 cm$^{-1}$, with a decreasing frequency trend as the number of L segments in the L layer increases. This is another evidence that the CC bonds belonging to the zig-zag chain bridge the chains in such a way that sizeable electrons conjugation between the diacetylene segments occurs.

As observed in 6-hT$^2$, aromatic systems are less prone to share $\pi$-electrons with the sp system, and this explains why the frequencies of the ECC bands of the GZY-II series are rather high and little sensitive to the width of the graphene-like ribbons.

As a last observation, we notice that the modulation of the ECC phonon frequency is remarkably stronger in GDY (dispersion of about 300 cm$^{-1}$) than in the whole GZY family (dispersion of 140 cm$^{-1}$).

In Figure 7 we illustrate the correlation between the frequency of the ECC line and the crystal relative energy $\Delta$E defined above. Similar to the case of the GDY family, we find a decreasing trend of the ECC frequency as the crystal become less stable. Also in this case larger $\Delta$E values correspond to less stable sp chains, showing a more cumulene like character (more equalized CC bonds). Figure 7.a and Figure 7.c show the expected trend of the ECC frequency versus $\Delta$E and BLA.

Interestingly, also for the GDZ family there is an outlier, namely 6-h$^3$L, showing its ECC frequency (2224 cm$^{-1}$) rather low in comparison with its $\Delta$E. Moreover, also its BLA value does not correlates well with the energy value.

Table 4. Sketches of the vibrational eigenvectors associated to the Raman active (A$_g$) phonons of GZY-I group, in the spectral region above 2000 cm$^{-1}$. In the last row the corresponding vibrational wavenumbers are reported.

| 6-hL | 6-hL$^2$ | 6-hL$^3$ | 6-L |
|------|----------|----------|-----|

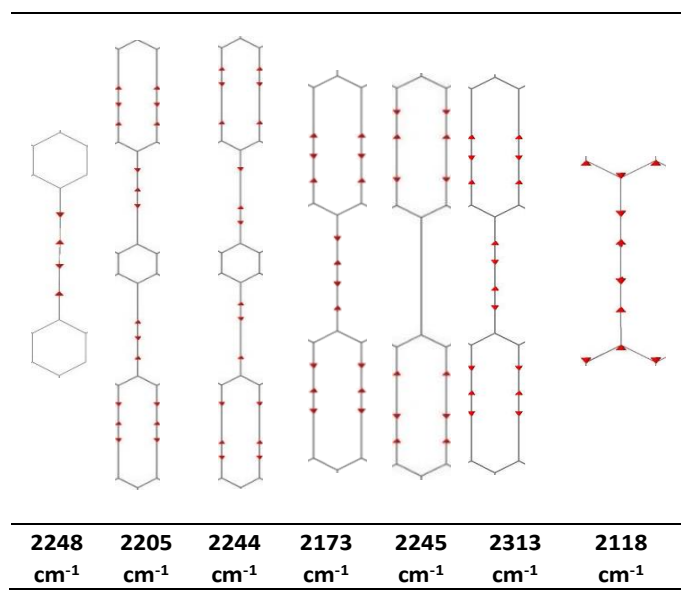

| 2248 cm$^{-1}$ | 2205 cm$^{-1}$ | 2244 cm$^{-1}$ | 2173 cm$^{-1}$ | 2245 cm$^{-1}$ | 2313 cm$^{-1}$ | 2118 cm$^{-1}$ |
|---|---|---|---|---|---|---|

A possible rationalization of such behaviour should consider the role of the sp$^2$ phase, which also contributes to $\Delta E$: notice that for 6-h$^3$L the sp/sp$^2$ ratio is the lowest in the series (0.5).

IR spectra. Above 2000 cm$^{-1}$ the GZY crystals show IR active B$_{3u}$ transitions characterized by dipole moment oscillations along the direction of the sp chains (x axis). Similar to their Raman counterpart, the number of IR active transitions equals the number of L units in the primitive cell. However, the IR spectrum shows only one band in most cases because of the very low IR intensity of the other transitions. Only in 6-hL$^2$ a satellite band can be observed in the IR spectrum, at a higher frequency than the main line (Figure 6).

The stronger IR line in this region corresponds, for all the GZY crystals, to the out of phase stretching (T$_i^-$) of the triple CC bonds of the chain. In cases with more than one chain per unit cell, it is described by the in-phase combination of T$_i^-$ of all the chains (see the vibrational eigenvectors sketched in Table 5). Interestingly, by looking at the data reported in Table 2, we can realize that phonons corresponding to similar displacements of the sp carbon can have very different IR intensities in the different crystals. In particular, 6-hL$^3$ shows a remarkably high IR intensity of the band at 2256 cm$^{-1}$ which could be ascribed to relevant charge fluxes associated with the collective vibration of the three L segments forming the L$^3$ layer. This last observation can be taken as further evidence of the high delocalization of $\pi$ electrons along the sequence made by three diacetylene segments.

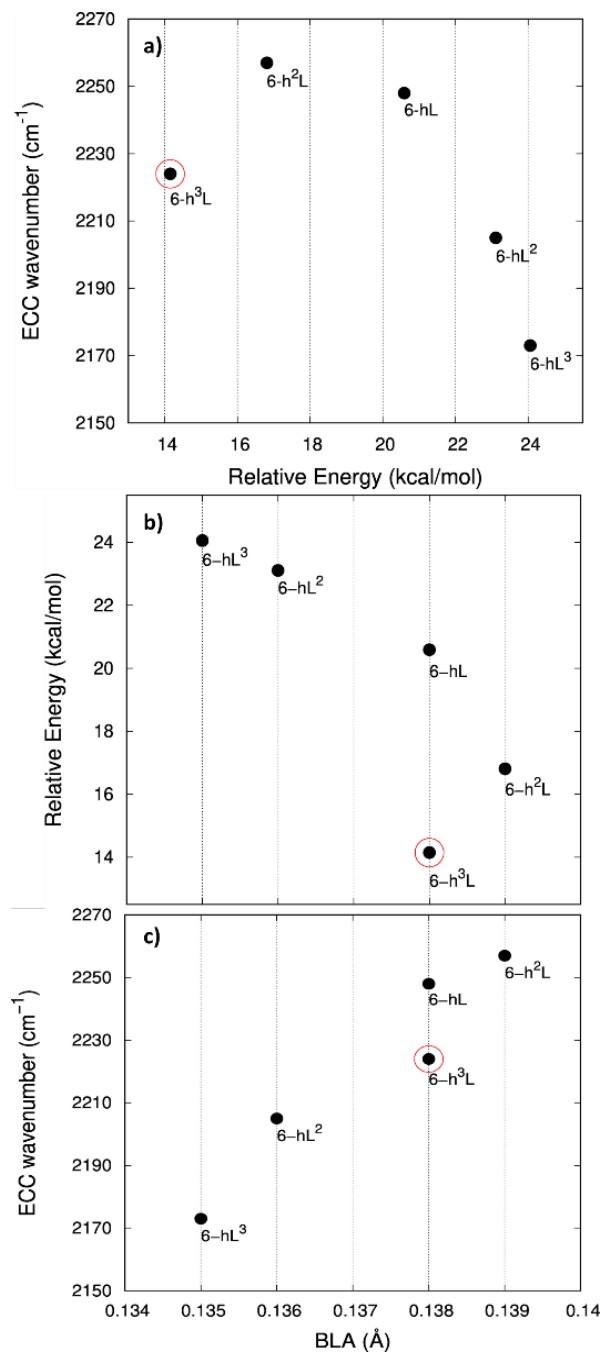

Figure 7. Correlation between physical quantities sensitive to $\pi$-electrons conjugation. Panel a: ECC phonon wavenumber vs. $\Delta E$ ($\Delta E$ is the relative crystal energy - per atom - referred to graphene, see text). Panel b: $\Delta E$ vs. BLA parameter. Panel c: ECC phonon wavenumber vs BLA. The data corresponding to 6-h$^3$L crystal are highlighted by a red circle.

Table 5. Sketches of the vibrational eigenvectors associated to the strongest IR active (B$_{3u}$) phonons of GZY-I, in the spectral region above 2000 cm$^{-1}$. In the last row the corresponding vibrational frequencies (cm$^{-1}$) are reported.

| 6-hL | 6-hL$^2$ | 6-hL$^3$ | 6-L |
|---|---|---|---|
| 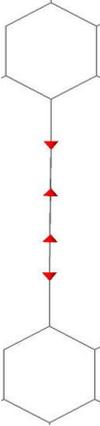 | 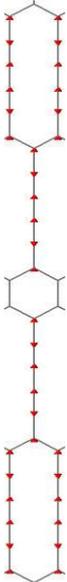 | 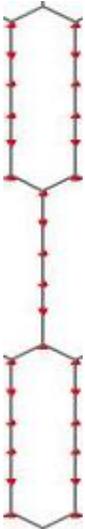 | 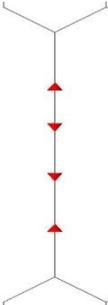 |
| 2255 cm$^{-1}$ | 2255 cm$^{-1}$ | 2256 cm$^{-1}$ | 2260 cm$^{-1}$ |

## Conclusions

The possibility to study how the properties of two-dimensional sp–sp$^2$ carbon materials are affected by the topology and connectivity of sp and sp$^2$ domains, or by their relative ratio is appealing for both fundamental and applied research with possible outcomes in technology. Despite the considerable theoretical efforts that have been spent on the assessment of the properties of these materials, most of the current literature focused on the γ polymorph of GDY, or the α- and β-GDY. Indeed, vibrational properties were analysed only for the γ polymorph [37] and for its nanoribbons [31] whereas a wide range of possible sp–sp$^2$ carbon materials, in the form of 2D crystals, could constitute interesting alternatives to γ-GDY.

We proposed here a rigorous analysis of the spectroscopic response of different GDY-based 2D crystals. Starting from a previously defined set of structures [58], we have investigated how the Raman active bands (ECC) are affected by different topologies and by the different nature of the sp and sp$^2$ domains. At the same time, the analysis of the IR spectra gave us the possibility to identify peculiar marker bands able to differentiate between different systems. We started the analysis with the most representative crystals. For these materials, we found a strong dependence of the ECC vibrational frequency on structure, topology, and stability. Indeed, the nature of the sp$^2$ domains strongly influences how the sp linear chains conjugate with them, strongly modulating the vibrational frequency associated with the ECC mode. Indeed, 6-H presented the most red-shifted ECC frequency while 6-hT$^2$ showed the most blue-shifted one. This result is supported by the very nice inter-correlation found between ECC, stability (ΔE), and BLA, which can describe all the physicochemical effects that take place in the analysed 2D crystals.

Similar considerations can be made for the second family (named here GZY), characterized by the presence of both h and L units. Interestingly, by looking at the calculated Raman spectra we can see that the GZY-II series shows a little dispersion of the ECC line, whereas for the GZY-I series the ECC – like phonon shows a dispersion of 130 cm$^{-1}$, with a decreasing frequency trend as the number of L segments in the L layer increases. This is another proof of the sizeable π-conjugation effects between the diacetylene segments. Also in this case, the observed trend is supported by the plots correlating ECC, ΔE, and BLA. Moreover, the modulation of the ECC phonon frequency is remarkably stronger in GDY (dispersion of about 300 cm$^{-1}$) than in the whole GZY family (dispersion of 140 cm$^{-1}$).

The recent advances in the experimental synthesis of GDYs make our results useful in the analysis of vibrational spectroscopy results and in evidencing specific structures. Indeed, for a few GDY systems (e.g. β-GDY, here called 6-HT$^2$, γ-GDY, here called 6-hT$^2$, and 6-R), experimental Raman spectra are available [68-70]. The experimental spectra are commonly characterized by wider peaks or bands. As expected, a peak in the high-frequency region is present at about 2100 cm$^{-1}$. In the region below 1600 cm$^{-1}$, a broad band centered at about 1500 cm$^{-1}$ is present in multilayer samples, whereas in ultrathin sheets it is structured with the presence of multiple peaks superimposed on the same band. Such broadening can be due to the presence of defects and confined crystal domains producing broadening of the Raman peaks due to confinement effects. The possible presence of disordered sp$^2$ carbon can also justify the typical shape of that band resembling the G and D peaks of amorphous carbon. A Raman peak at about 1900 cm$^{-1}$ is present in almost all the reported experimental spectra of GDY which is not reproduced by our calculations. This can be due to the presence of metal substrate used in the synthesis process and even metal atoms still present in the material. Indeed we have shown that the interaction with metallic substrate plays a relevant role in modifying the electronic and vibrational features. Possible strain and charge transfer effects, not considered in the freestanding crystals can substantially modify the Raman response giving rise to additional features in the region between 1600 and 2100 cm$^{-1}$ [71].

Our findings are relevant to the possibility to discriminate systems having different structural features, suggesting the application of vibrational spectroscopy for the characterization of topology. Moreover, these results shed light on a very peculiar electronic interplay between sp and sp$^2$ domains that are useful for the proper development of new nanostructured semiconductive sp–sp$^2$ carbon materials with tailored properties.

## Author Contributions

**Patrick Serafini:** Formal Analysis; Investigation; Writing – original draft; Data Curation; Software
**Alberto Milani**: Methodology; Software; Data curation; Conceptualization.
**Matteo Tommasini:** Writing – review & editing; Software; Validation; Data curation
**Chiara Castiglioni:** Conceptualization; Formal Analysis; Validation; Methodology; Writing – review & editing
**Davide M. Proserpio:** Conceptualization; Writing – review & editing
**Carlo Bottani:** Writing – review & editing
**Carlo Casari:** Supervision; Visualization; Validation; Writing – review & editing; Project administration

## Conflicts of interest

There are no conflicts to declare.

## Acknowledgements

P.S., A.M., C.C., M.T., C.E.B. and C.S.C. acknowledge funding from the European Research Council (ERC) under the European Union's Horizon 2020 research and innovation program ERC Consolidator Grant (ERC CoG2016 EspLORE grant agreement no. 724610, website: www.esplore.polimi.it).

## Notes and references